\documentclass[aps,prl,preprint,superscriptaddress,showpacs,floatfix]{revtex4-1}

\usepackage{graphicx}
\usepackage{dcolumn}
\newcolumntype{d}{D{.}{.}{2.5}}
\newcolumntype{s}{D{.}{.}{1.2}}

\begin{document}

\title{Finding the Rashba-type spin-splitting from interband scattering in quasiparticle interference maps}

\author{Manuel Steinbrecher}
\affiliation{Physikalisches Institut and Center for Nanotechnology
(CeNTech), Westf\"alische Wilhelms-Universit\"at M\"unster, 48149
M\"unster, Germany}

\author{Hasmik Harutyunyan}
\affiliation{Physikalisches Institut and Center for Nanotechnology
(CeNTech), Westf\"alische Wilhelms-Universit\"at M\"unster, 48149
M\"unster, Germany}

\author{Christian R. Ast}
\affiliation{Max-Planck-Institut f\"ur Festk\"orperforschung,
70569 Stuttgart, Germany}

\author{Daniel Wegner}
\affiliation{Physikalisches Institut and Center for Nanotechnology
(CeNTech), Westf\"alische Wilhelms-Universit\"at M\"unster, 48149
M\"unster, Germany}

\date{\today}

\begin{abstract}
We have studied the BiCu$_2$/Cu(111) surface alloy using
low-temperature scanning tunneling microscopy and spectroscopy. We
observed standing waves caused by scattering off defects and step
edges. Different from previous studies on similar Rashba-type
surfaces, we identified multiple scattering vectors that originate
from various intraband as well as interband scattering processes. A
detailed energy-dependent analysis of the standing-wave patterns
enables a quantitative determination of band dispersions, including
the Rashba splitting. The results are in good agreement with ARPES
data and demonstrate the usefulness of this strategy to determine
the band structure of Rashba systems. The lack of other possible
scattering channels will be discussed in terms of spin conservation
and hybridization effects. The results open new possibilities to
study spin-dependent scattering on complex spin-orbit coupled
surfaces.
\end{abstract}

\pacs{68.37.Ef, 73.20.At, 71.70.Ej, 72.10.Fk}

\maketitle

%
%

The Bychkov-Rashba effect lifts the spin-degeneracy at surfaces and
interfaces in environments with a sizeable spin-orbit interaction
\cite{bychkov_properties_1984}. Up to date, many different systems
with various kinds of splitting strengths have been identified
\cite{lashell_spin_1996,koroteev_strong_2004,AstPRL2007}. Its
characteristic band dispersion is readily recognizable in angular
resolved photoemission experiments and generally good agreement is
achieved between theoretical calculations and experimentally
observed electronic structures
\cite{Petersen00,AstPRL2007,Bihlmayer2007,Gierz2010,Moreschini2009,
BentmannEPL2009,Bentmann2011,Uenal2012}. The Rashba parameter can
also be extracted from the local density of states measured by
scanning tunneling spectroscopy (STS) \cite{Ast2007}. However,
identifying the Rashba-type spin-splitting from quasiparticle
interference (QPI) patterns measured by STS remains elusive. The
reason for this is the defined spin-polarization for each singly
degenerate state resulting in forbidden backscattering. As a
consequence, the Rashba signature is suppressed in QPI patterns of
isotropic band structures, making the pattern look like it
originates from a spin-degenerate state \cite{Petersen00}. In highly
anisotropic surface states, e.g.\ Bi(110) \cite{Pascual04} or
Bi$_{1-x}$Sb$_x$(111) \cite{roushan_topological_2009}, the
``spin-selection rule'' strongly reduces the set of possible
momentum transfers, so that the presence of a Rashba-type
spin-splitting can be observed qualitatively. QPI patterns from
multiple scattering events can also give a qualitative difference
between spin-split states and spin-degenerate states
\cite{walls_spin-orbit_2007}. However, a quantitative statement
about the strength of the spin-splitting from the analysis of QPI
patterns has not been reported so far.

Here, we show that it is possible to extract the Rashba parameter
from interband scattering events in QPI patterns using scanning
tunneling microscopy (STM) and STS. For this proof of principle, we
choose the well-known system BiCu$_2$/Cu(111). Differential
conductance maps at different energies reveal standing-wave patterns
due to scattering at defects and step edges. A careful
Fourier-transform (FT) analysis reveals several scattering vectors,
which can be related to intraband transitions within the first
$sp_z$-type surface state, intraband transitions within the second
($p_{xy}$) surface state, and interband transitions between the
first and second surface state. A detailed data set at several
energies is used to recover the dispersion relations of both surface
states, including the Rashba-type spin-splitting.

The experiments were performed in ultrahigh vacuum (UHV) using a
commercial low-temperature STM (Createc LT-STM) operated at $T = 6\,
\text{K}$. A Cu(111) single-crystal substrate was first cleaned by
standard sputter-annealing procedures. The BiCu$_2$ surface alloy
was then grown by evaporating one third of a monolayer of Bi onto
the Cu substrate from a Knudsen cell. During and for 10 minutes
after the deposition, the substrate was held at about 390~K. The
sample was then transferred \emph{in situ} into the cryogenic STM.
Topography images were taken in constant-current mode. STS was
performed by measuring the differential conductance $dI/dV$ as a
function of the sample bias $V$ by standard lock-in techniques
($V_{\text{mod}} = 1-20\, \text{mV}$ (rms), frequency $768.3\,
\text{Hz}$) under open-feedback conditions. $dI/dV$ maps at various
energies were taken in constant-current as well as constant-height
mode using $V_{\text{mod}} = 20\, \text{mV}$. The maps were then
Fourier-transformed and analyzed using a commercial image analysis
software (Image Metrology SPIP).

%
%

\begin{figure}
\begin{center}
\includegraphics[width=0.4\textwidth]{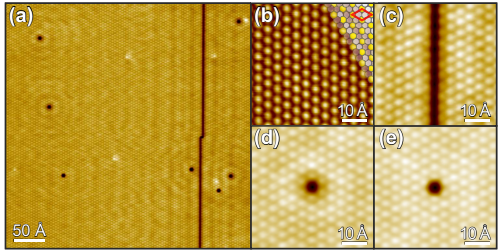}
\caption{\label{fig1} (Color online) STM topography of
BiCu$_2$/Cu(111). (a) The overview image shows a smooth surface with
an antiphase boundary and a few point defects. Due to the low sample
bias (1~mV, 100~pA), standing waves can be observed. (b) Close-up
view of the $\sqrt{3} \times \sqrt{3}$ structure (1~mV, 63~nA). Bi
atoms appear bright (cf.\ structural model on top right). (c) A
close-up view of the antiphase boundary. (d,e) Zoom images of the
top left and bottom left point defects in (a) show small
topographical differences.}
\end{center}
\end{figure}

\begin{figure}
\begin{center}
\includegraphics[width=0.45\textwidth]{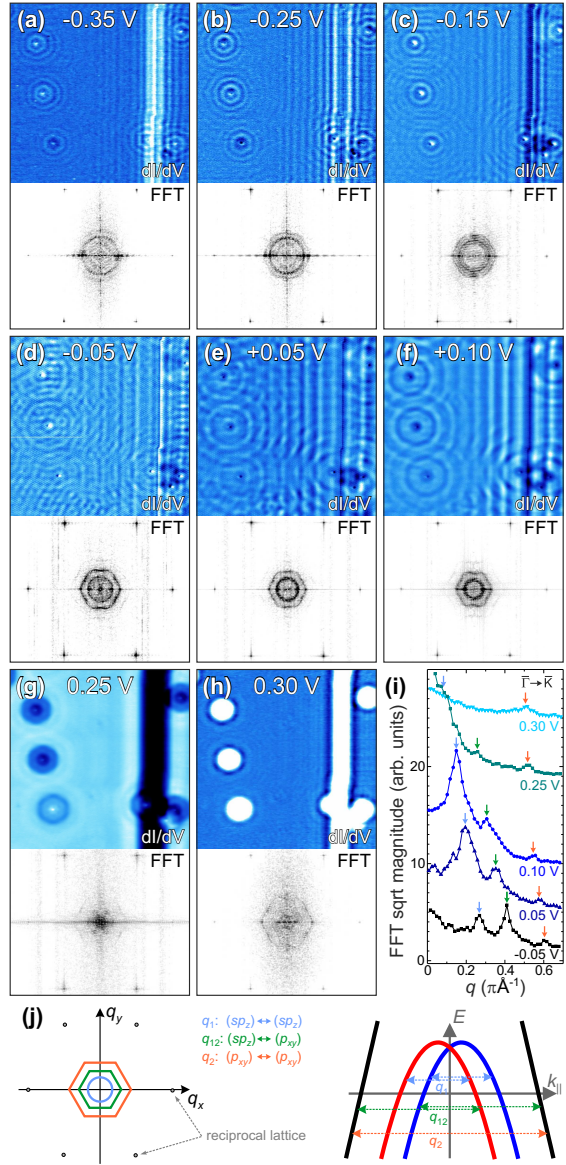}
\caption{\label{fig2} (Color online) FT-STS analysis of
BiCu$_2$/Cu(111). (a-h)~$dI/dV$ maps and corresponding FT images at
different sample voltages. (i) FT cross sections along a
$\overline{\Gamma}\overline{\text{K}}$ direction show the
bias-dependent trend of the scattering features. (j)~Schematic
representation of the (up to) four features visible in the FT images
and proposed assignments deduced from the band structure.}
\end{center}
\end{figure}

Figure~\ref{fig1} shows STM images of the surface alloy. We found
large atomically smooth surfaces with typical sizes of about $1000
\times 1000$ \AA$^2$, limited by step edges and antiphase
boundaries. Fig.~\ref{fig1}a shows an overview of a large terrace. A
close-up view (b) of the surface reveals the well-known BiCu$_2$
surface with a $\sqrt{3} \times \sqrt{3}$ structure and a lattice
constant of 4.42~\AA\ (cf.\ structural model superimposed in (b))
\cite{Kaminski2005}. Bi atoms appear as bright protrusions, while
the surrounding Cu atoms cannot be resolved atomically
\cite{Ast2007,Gierz2010}. On the right-hand side, the overview image
shows a trench. As can be seen in (c), this feature corresponds to
an antiphase boundary of two superstructures on the left and
right-hand sides, respectively. In addition, point defects can be
observed, which are at positions where a Bi atom should be located.
We assume that they are sites where a Cu atom has not been replaced
by a Bi atom within the surface layer. We have observed two types of
point scatterers that appear slightly different in topography images
at low bias (d and e). While the reason is unknown, both show
qualitatively identical scattering behavior (cf. Fig.~\ref{fig2})
and will therefore be treated as being the same. The average defect
density is less than $1/1000^{\text{th}}$ monolayer ($\approx 9
\cdot 10^{-5} \text{\AA}^{-2}$, i.e., defect separation $\approx
120$~\AA). As we will discuss later, this corresponds to a very good
sample quality.

The topography image in Fig.~\ref{fig1}a already shows standing
waves on the surface: a planar wave front is scattered off the
antiphase boundary, and the point defects are surrounded by circular
waves. As is well-known, the energy-dependent wavelengths of these
standing waves are directly connected to the dispersion relation of
the surface bands \cite{Crommie1993,Hasegawa1993}. To enable a
thorough analysis of the underlying scattering processes, we have
used the FT-STS method at various energies \cite{Sprunger1997}.
Fig.~\ref{fig2} summarizes some of the acquired $dI/dV$ maps and
corresponding 2D-Fourier transformations \footnote{See Supplemental
Material at [URL] for a movie of $dI/dV$ maps between -0.45~V and
+0.4~V.}. Interestingly, the maps in Fig.~\ref{fig2} undoubtedly
show that the standing-wave pattern is caused by more than one
wavelength.

In the FT images, up to four features can be identified that are
schematically depicted in Fig.~\ref{fig2}j. The strongest intensity
is found in six spots with hexagonal symmetry (black dots in (j);
they reflect the reciprocal lattice of the $\sqrt{3} \times
\sqrt{3}$ surface reconstruction and allow us to identify the
$\overline{\Gamma}\overline{\text{M}}$ directions at angles
0$^{\circ}$ (horizontal axis), 60$^{\circ}$ and 120$^{\circ}$,
respectively, while the $\overline{\Gamma}\overline{\text{K}}$
directions are found at 30$^{\circ}$, 90$^{\circ}$, and
150$^{\circ}$. The second strongest FT signal is a small circular
ring (light blue circle in (j)) whose radius decreases with
increasing sample bias. At $E-E_F = 0.25$~eV (g), this feature is
still visible as a narrow signal in the center of the FT image
\footnote{Determining $q_1$ and $q_{12}$ close to the band maximum
from FT cross sections becomes difficult; here our analysis is based
on a combined FT and $dI/dV$-linescan analysis.}, but disappears
above 0.25~eV (h). In addition, a second ring is clearly visible
that has a hexagonal shape (green in (j)). This feature also becomes
smaller as we go from $-0.3$~eV to $0.25$~eV and vanishes at higher
energies (g,h). The final feature is a faint third ring that also
has hexagonal shape (orange in (j)). Although the intensity is
rather weak, a careful analysis via cross sections of the FT images
(cf.\ orange markers in Fig.~\ref{fig2}i) allows a clear
determination of this scattering feature between $-0.25$~eV and
$0.5$~eV.

A comparison of the distinct energy and angular dependence of the
three scattering rings with band-structure calculations and ARPES
measurements permits an unambiguous assignment of the features to
specific scattering transitions in the band structure. The smallest
(light blue) feature is isotropic and dominates the FT images up to
0.25~eV but suddenly vanishes for higher energies
(Fig.~\ref{fig2}i). This is the expected behavior for scattering
within the first $sp_z$-type surface state
\cite{Moreschini2009,BentmannEPL2009}. As schematically shown in
Fig.~\ref{fig2}j, this band is Rashba-split: for any
$E(k_{\parallel})$ plane through the $\overline{\Gamma}$ point, two
parabolas of opposite spin exist that are shifted away from the
$\overline{\Gamma}$-point by a wave-vector offset $k_0$. During a
scattering event, the spin is conserved
\cite{Petersen00,Pascual04,Hirayama2011}. Therefore,
$sp_z$-intraband scattering only occurs between the branches of the
same spin, leading to only one possible scattering vector $q_1$, as
depicted in Fig.~\ref{fig2}j. We note that previous studies tend to
interpret the scattering vector $q$ as identical to the wave vector
$k_{\parallel}$ in the energy dispersion through the relation $q/2 =
k_{\parallel}$ \cite{Crommie1993,Hasegawa1993,Hirayama2011}. This
assignment is valid whenever a dispersion is symmetric around
$\overline{\Gamma}$ \cite{Sprunger1997,Pascual2001,Schouteden2009}.
For the Rashba-split parabolas, the scattering can only give
information on the effective mass and maximum of the band, but
information on the wave vector offset $k_0$ in $k_{\parallel}$
direction (i.e., the Rashba splitting) is lost \cite{Petersen00}.

\begin{figure}
\begin{center}
\includegraphics[width=0.45\textwidth]{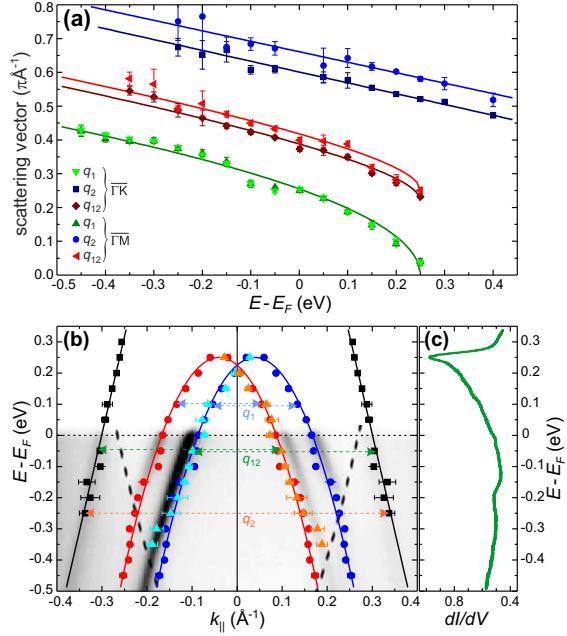}
\caption{\label{fig3} (Color online) Analysis of scattering channels
and corresponding electronic structure. (a) Energy-dependence of
$q_1$, $q_2$, and $q_{12}$ for $\overline{\Gamma}\overline{M}$ and
$\overline{\Gamma}\overline{K}$ directions, respectively. Solid
lines are fits to the data. (b) Band dispersions along
$\overline{\Gamma}\overline{M}$ (solid lines) as determined from the
fit. The points are experimentally derived $k$ values calculated
from the $q_i$ vectors as indicated by the arrows (cf.\
Fig.~\ref{fig2}j). For direct comparison, an ARPES result of the
band dispersion is also shown (reproduced from
Ref.~\onlinecite{BentmannEPL2009}). (c) The STS spectrum exhibits a
peak at 0.25~eV above $E_F$ that corresponds to the $sp_z$-band
maximum, in agreement with the fit in (b).}
\end{center}
\end{figure}

In order to identify the outer (orange) scattering ring, its
distinct hexagonal anisotropy can be compared to ARPES and DFT
results \cite{Moreschini2009,BentmannEPL2009}. A hexagonally shaped
constant-energy surface is expected for the second surface band
exhibiting $p_{xy}$ character. The wave vector $k_{\parallel}$ is
larger along $\overline{\Gamma}\overline{\text{M}}$ compared to the
$\overline{\Gamma}\overline{\text{K}}$ direction in accordance with
our observed anisotropy. Also, the projected bulk-band edge, which
is folded back due to the surface reconstruction, exhibits a
hexagonal shape close to $\overline{\Gamma}$
\cite{Moreschini2009,BentmannEPL2009,Uenal2012}. Here, however, the
maximum wave vector is along $\overline{\Gamma}\overline{\text{K}}$.
Hence, we can rule out transitions from or to bulk states. In fact,
we find that the magnitude of the scattering vector is in very good
agreement with the expected value for intraband scattering within
the second surface state ($q_2$ in Fig.~\ref{fig2}j). This also
explains, why we can observe this feature in $dI/dV$ maps far beyond
0.25~eV (g-i). Within the energy range presented here, this band is
spin-degenerate and symmetric around $\overline{\Gamma}$. Thus, we
can directly deduce the dispersion relation via
$k_{\parallel}(p_{xy}) = q_2/2$.

The anisotropy of the middle (green) ring seen in the FT images is
identical to the shape of the outer ring associated with $q_2$.
Further, as the feature vanishes above 0.25~eV, the transition
likely also involves the $sp_z$-band. Ruling out scattering channels
that involve bulk states, the only possibility for this transition
to occur is due to interband scattering between the $p_{xy}$- and
$sp_z$-surface states. As the latter is Rashba-split, two different
transitions can in principle be expected involving either the inner
or the outer branch of the $sp_z$-band. A comparison of the
magnitude reveals that our feature corresponds to scattering from
(or to) the inner branch (denoted $q_{12}$ in Fig.~\ref{fig2}j).
Surprisingly, we have not found any indication for scattering
between the the $p_{xy}$-band and the outer $sp_z$-branch (cf.\
discussion below).

Fig.~\ref{fig3}a summarizes the energy dependence of the magnitude
of all three scattering processes $q_1$, $q_2$, and $q_{12}$ along
$\overline{\Gamma}\overline{\text{M}}$ and
$\overline{\Gamma}\overline{\text{K}}$ as determined from the FT-STS
data. In order to determine the band structure parameters from this
data set, we assume a parabolic dispersion with a Rashba-type
spin-splitting for the $sp_z$-band as well as a spin-degenerate
linear dispersion for the $p_{xy}$-band. The linear dispersion is
justified because in the relevant energy interval the band is
located between the high-symmetry points in the vicinity of the
inflection point, so that the band curvature can be neglected.
Solving for the wave vector, the dispersion is
$k(E) = \mu[2m^*(E-E_{0})/\hbar^2]^{1/2}+\lambda k_0$
with $\mu,\,\lambda=\pm 1$, where $\mu$ denotes the parabola branch
and $\lambda$ denotes the Rashba branch
\cite{Moreschini2009,Mirhosseini2009,BentmannEPL2009,Bentmann2011,Uenal2012}.
Here, $m^*$ is the effective mass and $E_{0}$ is the energy offset
of the $sp_z$-band. Combining the intraband transitions $q_1$ and
$q_2$ with the interband transition $q_{12}$, we can extract the
wave vector offset determining the Rashba-type spin-splitting:
$k_0=(q_1 +q_2)/2-q_{12}$.
In order to decrease uncertainties of the fit procedure, we have
simultaneously fit all three scattering vectors in both
$\overline{\Gamma}\overline{\text{M}}$ and
$\overline{\Gamma}\overline{\text{K}}$ directions (i.e., six data
sets) using the same fit parameters (seven fit parameters: four
describing the $p_{xy}$-band in
$\overline{\Gamma}\overline{\text{M}}$ and
$\overline{\Gamma}\overline{\text{K}}$ as well as $m^*$, $E_0$, and
$k_0$). The solid lines in Fig.~\ref{fig3}a are the resulting fit
curves. Overall, we find very good agreement between the raw data
and the fit. Table~\ref{table1} summarizes the corresponding fit
parameters in comparison with literature values. Using these values,
we can plot the dispersion relations of the $p_{xy}$-band as well as
the Rashba-split $sp_z$-band. The result is shown in
Fig.~\ref{fig3}b. The fit parameters and the extracted band
dispersions agree well with other published data (see Table
\ref{table1}), despite the simplicity of the model used in this
analysis. This demonstrates that we are able to fully recover the
surface bands of this Rashba system, including the Rashba splitting
$k_0$, the Rashba parameter $\alpha_R = \hbar^2 k_0/m^*$, and the
Rashba energy $E_R= \hbar^2 k_0^2/2m^*$.

We find a slightly higher band maximum $E_0$ compared to previous
studies. This is also confirmed by local STS spectra in
Fig.~\ref{fig3}c, where the band maximum appears as an asymmetric
sharp peak at about 250~meV due to a singularity in the density of
states \cite{Ast2007, Wegner2006}. We note that on other samples
with higher local defect concentrations, we found reduced values for
$E_0$. Moreover, confinement effects were evident on narrow
terraces: as expected for a downward dispersing band, the peak in
STS spectra was shifted to lower energies, roughly following an
$L^{-2}$ dependence (where $L$ is the terrace width). As
photoemission experiments probe large sample areas, we suggest that
the previously reported smaller $E_0$ values may be a consequence of
small terraces and/or reduced sample quality.

The magnitude of $m^*$ in our study is a bit smaller than reported
before. However, our simple fit model merely considers a parabolic
shape of the $sp_z$-band. While this is a good assumption for
energies above $-0.2$~eV, ARPES measurements found a significant
deviation for the outer branches: due to strong hybridization with
bulk states below $-0.2$~eV, the dispersion bends down and exhibits
a much steeper slope \cite{Moreschini2009,BentmannEPL2009}. In our
model we only use one average effective mass, therefore we expect it
to be smaller.

\begingroup
\squeezetable
\begin{table}
\caption{Results of parameters for the Rashba-split $sp_z$-type
surface state of this work (FT-STS) compared to values reported in
the literature.}
\begin{ruledtabular}
\begin{tabular}{dddddd} 
                                       & \multicolumn{1}{c}{$m^*$\,($m_e$)} &  \multicolumn{1}{c}{$E_0\,(\text{eV})$}  &  \multicolumn{1}{c}{$k_0\,(\text{\AA}^{-1})$}  &  \multicolumn{1}{c}{$\alpha_R\,(\text{eV{\AA}})$}  &  \multicolumn{1}{c}{$E_R\,(\text{eV})$}  \\
 \hline
 \multicolumn{1}{l}{FT-STS} &   -0.25(2)         &   0.25(1)           &   0.04(1)                &   1.2(3)                 &   0.02(1)      \\
 \multicolumn{1}{l}{Ref.~\onlinecite{Moreschini2009}} &   -0.27            &   0.23              &   0.03                   &   1.0                    &   0.015        \\
 \multicolumn{1}{l}{Ref.~\onlinecite{BentmannEPL2009}} &   -0.30            &   0.231             &   0.032                  &   0.82                   &   0.013        \\
 \multicolumn{1}{l}{Ref.~\onlinecite{Uenal2012}} &   -0.37            &  0.23  &   0.04                   &   0.83                   &   0.020        \\
\end{tabular}
\end{ruledtabular}
\label{table1}
\end{table}
\endgroup

In agreement with previous observations on Bi surfaces and alloys,
we found that the spin is conserved during the scattering process
\cite{Petersen00,Pascual04,Hirayama2011}. As a result, we only
observe scattering from the outer to the inner branch of the
$sp_z$-band and vice versa ($q_1$). The $p_{xy}$-band is
spin-degenerate in the energy region considered here
\cite{BentmannEPL2009,Mirhosseini2009,Bentmann2011}. Therefore, we
were able to observe intraband scattering ($q_2$) as well as
interband scattering to the inner branch of the $sp_z$-band
($q_{12}$). However, it is surprising at a first glance that we have
not observed a scattering process involving the $p_{xy}$-band and
the outer branch of the $sp_z$-band. While this transition is not
forbidden \emph{per se}, a bad wave-function overlap between initial
and final state may lead to a vanishing scattering amplitude.
Interestingly, ARPES and DFT results show that the inner and outer
$sp_z$-branches behave very differently
\cite{Moreschini2009,BentmannEPL2009}: when crossing the bulk-band
edge, the outer branch strongly hybridizes with the bulk states.
This is obvious from the broadening, the loss of spectral density
and the sudden deviation of the dispersion from the parabolic shape.
On the other hand, the inner branch is barely influenced by bulk
bands, as is the $p_{xy}$-band, indicating weak hybridization for
these states. While further investigations are required, preliminary
model calculations support the possibility that a reduced orbital
overlap between the outer $sp_z$-branch and the $p_{xy}$-band
combined with a large magnitude of the scattering vector can lead to
a significantly reduced scattering amplitude that can be hidden in
the background noise of our FT-STS data \cite{Krueger}.

In conclusion, we have studied the QPI patterns in $dI/dV$ maps of
the 2D-band structure in the BiCu$_2$/Cu(111) surface alloy using
STS. While the scattering vectors from intraband scattering in the
spin-split $sp_z$-band do not carry any information about the
Rashba-type spin-splitting, a combination of intraband and interband
scattering vectors from the $sp_z$-band and the spin-degenerate
$p_{xy}$-band readily reveals the wave vector offset $k_0$ of the
Rashba-type spin-splitting. Concerning topological insulators (TIs),
the Rashba constant is ill-defined in these systems due to the
different band topology. Nevertheless, our results show that
additional scattering channels relax the condition of ``forbidden
backscattering'', which also applies for the TIs. This means that a
TI can only feature forbidden backscattering if there is exactly
{\em one} singly degenerate band crossing the Fermi level. Our
results present a nice experimental proof of principle that the
information about the Rashba-type spin-splitting is contained in QPI
patterns and can be extracted when considering interband scattering.
In this respect, the BiCu$_2$/Cu(111) surface alloy as well as other
similar surface alloys can be used as excellent model systems to
study the effects of spin-dependent scattering from, e.g., magnetic
or non-magnetic adsorbates, multiple scattering events as well as
confinement effects from smaller terraces or islands.

%
%
\begin{acknowledgments}
D.\ W.\ and C.\ R.\ A.\ acknowledge funding from the
Emmy-Noether-Program of the Deutsche Forschungsgemeinschaft (DFG),
projects WE 4104/2-1 and AS 152/3-1, respectively. We thank Peter
Kr\"uger for stimulating discussions.
\end{acknowledgments}

\bibliography{Rashba-STM_v4}

\begin{thebibliography}{27}%
\makeatletter
\providecommand \@ifxundefined [1]{%
 \@ifx{#1\undefined}
}%
\providecommand \@ifnum [1]{%
 \ifnum #1\expandafter \@firstoftwo
 \else \expandafter \@secondoftwo
 \fi
}%
\providecommand \@ifx [1]{%
 \ifx #1\expandafter \@firstoftwo
 \else \expandafter \@secondoftwo
 \fi
}%
\providecommand \natexlab [1]{#1}%
\providecommand \enquote  [1]{``#1''}%
\providecommand \bibnamefont  [1]{#1}%
\providecommand \bibfnamefont [1]{#1}%
\providecommand \citenamefont [1]{#1}%
\providecommand \href@noop [0]{\@secondoftwo}%
\providecommand \href [0]{\begingroup \@sanitize@url \@href}%
\providecommand \@href[1]{\@@startlink{#1}\@@href}%
\providecommand \@@href[1]{\endgroup#1\@@endlink}%
\providecommand \@sanitize@url [0]{\catcode `\\12\catcode `\$12\catcode
  `\&12\catcode `\#12\catcode `\^12\catcode `\_12\catcode `\%12\relax}%
\providecommand \@@startlink[1]{}%
\providecommand \@@endlink[0]{}%
\providecommand \url  [0]{\begingroup\@sanitize@url \@url }%
\providecommand \@url [1]{\endgroup\@href {#1}{\urlprefix }}%
\providecommand \urlprefix  [0]{URL }%
\providecommand \Eprint [0]{\href }%
\providecommand \doibase [0]{http://dx.doi.org/}%
\providecommand \selectlanguage [0]{\@gobble}%
\providecommand \bibinfo  [0]{\@secondoftwo}%
\providecommand \bibfield  [0]{\@secondoftwo}%
\providecommand \translation [1]{[#1]}%
\providecommand \BibitemOpen [0]{}%
\providecommand \bibitemStop [0]{}%
\providecommand \bibitemNoStop [0]{.\EOS\space}%
\providecommand \EOS [0]{\spacefactor3000\relax}%
\providecommand \BibitemShut  [1]{\csname bibitem#1\endcsname}%
\let\auto@bib@innerbib\@empty
\bibitem [{\citenamefont {Bychkov}\ and\ \citenamefont
  {Rashba}(1984)}]{bychkov_properties_1984}%
  \BibitemOpen
  \bibfield  {author} {\bibinfo {author} {\bibfnamefont {Y.~A.}\ \bibnamefont
  {Bychkov}}\ and\ \bibinfo {author} {\bibfnamefont {E.~I.}\ \bibnamefont
  {Rashba}},\ }\href
  {http://apps.isiknowledge.com/full_record.do?product=WOS&search_mode=GeneralSearch&qid=1&SID=Y17Nl5Aa8DnFPGGOChE&page=1&doc=9}
  {\bibfield  {journal} {\bibinfo  {journal} {JETP Lett.}\ }\textbf {\bibinfo
  {volume} {39}},\ \bibinfo {pages} {78} (\bibinfo {year} {1984})}\BibitemShut
  {NoStop}%
\bibitem [{\citenamefont {{LaShell}}\ \emph {et~al.}(1996)\citenamefont
  {{LaShell}}, \citenamefont {{McDougall}},\ and\ \citenamefont
  {Jensen}}]{lashell_spin_1996}%
  \BibitemOpen
  \bibfield  {author} {\bibinfo {author} {\bibfnamefont {S.}~\bibnamefont
  {{LaShell}}}, \bibinfo {author} {\bibfnamefont {B.~A.}\ \bibnamefont
  {{McDougall}}}, \ and\ \bibinfo {author} {\bibfnamefont {E.}~\bibnamefont
  {Jensen}},\ }\href {\doibase 10.1103/PhysRevLett.77.3419} {\bibfield
  {journal} {\bibinfo  {journal} {Phys. Rev. Lett.}\ }\textbf {\bibinfo
  {volume} {77}},\ \bibinfo {pages} {3419} (\bibinfo {year}
  {1996})}\BibitemShut {NoStop}%
\bibitem [{\citenamefont {Koroteev}\ \emph {et~al.}(2004)\citenamefont
  {Koroteev}, \citenamefont {Bihlmayer}, \citenamefont {Gayone}, \citenamefont
  {Chulkov}, \citenamefont {Bl\"{u}gel}, \citenamefont {Echenique},\ and\
  \citenamefont {Hofmann}}]{koroteev_strong_2004}%
  \BibitemOpen
  \bibfield  {author} {\bibinfo {author} {\bibfnamefont {Y.~M.}\ \bibnamefont
  {Koroteev}}, \bibinfo {author} {\bibfnamefont {G.}~\bibnamefont {Bihlmayer}},
  \bibinfo {author} {\bibfnamefont {J.~E.}\ \bibnamefont {Gayone}}, \bibinfo
  {author} {\bibfnamefont {E.~V.}\ \bibnamefont {Chulkov}}, \bibinfo {author}
  {\bibfnamefont {S.}~\bibnamefont {Bl\"{u}gel}}, \bibinfo {author}
  {\bibfnamefont {P.~M.}\ \bibnamefont {Echenique}}, \ and\ \bibinfo {author}
  {\bibfnamefont {P.}~\bibnamefont {Hofmann}},\ }\href {\doibase
  10.1103/PhysRevLett.93.046403} {\bibfield  {journal} {\bibinfo  {journal}
  {Phys. Rev. Lett.}\ }\textbf {\bibinfo {volume} {93}},\ \bibinfo {pages}
  {046403} (\bibinfo {year} {2004})}\BibitemShut {NoStop}%
\bibitem [{\citenamefont {Ast}\ \emph {et~al.}(2007{\natexlab{a}})\citenamefont
  {Ast}, \citenamefont {Henk}, \citenamefont {Ernst}, \citenamefont
  {Moreschini}, \citenamefont {Falub}, \citenamefont {Pacil\'e}, \citenamefont
  {Bruno}, \citenamefont {Kern},\ and\ \citenamefont {Grioni}}]{AstPRL2007}%
  \BibitemOpen
  \bibfield  {author} {\bibinfo {author} {\bibfnamefont {C.~R.}\ \bibnamefont
  {Ast}}, \bibinfo {author} {\bibfnamefont {J.}~\bibnamefont {Henk}}, \bibinfo
  {author} {\bibfnamefont {A.}~\bibnamefont {Ernst}}, \bibinfo {author}
  {\bibfnamefont {L.}~\bibnamefont {Moreschini}}, \bibinfo {author}
  {\bibfnamefont {M.~C.}\ \bibnamefont {Falub}}, \bibinfo {author}
  {\bibfnamefont {D.}~\bibnamefont {Pacil\'e}}, \bibinfo {author}
  {\bibfnamefont {P.}~\bibnamefont {Bruno}}, \bibinfo {author} {\bibfnamefont
  {K.}~\bibnamefont {Kern}}, \ and\ \bibinfo {author} {\bibfnamefont
  {M.}~\bibnamefont {Grioni}},\ }\href {\doibase 10.1103/PhysRevLett.98.186807}
  {\bibfield  {journal} {\bibinfo  {journal} {Phys. Rev. Lett.}\ }\textbf
  {\bibinfo {volume} {98}},\ \bibinfo {pages} {186807} (\bibinfo {year}
  {2007}{\natexlab{a}})}\BibitemShut {NoStop}%
\bibitem [{\citenamefont {Petersen}\ and\ \citenamefont
  {Hedeg{\aa}rd}(2000)}]{Petersen00}%
  \BibitemOpen
  \bibfield  {author} {\bibinfo {author} {\bibfnamefont {L.}~\bibnamefont
  {Petersen}}\ and\ \bibinfo {author} {\bibfnamefont {P.}~\bibnamefont
  {Hedeg{\aa}rd}},\ }\href@noop {} {\bibfield  {journal} {\bibinfo  {journal}
  {Surf. Sci.}\ }\textbf {\bibinfo {volume} {459}},\ \bibinfo {pages} {49}
  (\bibinfo {year} {2000})}\BibitemShut {NoStop}%
\bibitem [{\citenamefont {Bihlmayer}\ \emph {et~al.}(2007)\citenamefont
  {Bihlmayer}, \citenamefont {Bl\"ugel},\ and\ \citenamefont
  {Chulkov}}]{Bihlmayer2007}%
  \BibitemOpen
  \bibfield  {author} {\bibinfo {author} {\bibfnamefont {G.}~\bibnamefont
  {Bihlmayer}}, \bibinfo {author} {\bibfnamefont {S.}~\bibnamefont {Bl\"ugel}},
  \ and\ \bibinfo {author} {\bibfnamefont {E.~V.}\ \bibnamefont {Chulkov}},\
  }\href {\doibase 10.1103/PhysRevB.75.195414} {\bibfield  {journal} {\bibinfo
  {journal} {Phys. Rev. B}\ }\textbf {\bibinfo {volume} {75}},\ \bibinfo
  {pages} {195414} (\bibinfo {year} {2007})}\BibitemShut {NoStop}%
\bibitem [{\citenamefont {Gierz}\ \emph {et~al.}(2010)\citenamefont {Gierz},
  \citenamefont {Stadtm\"uller}, \citenamefont {Vuorinen}, \citenamefont
  {Lindroos}, \citenamefont {Meier}, \citenamefont {Dil}, \citenamefont
  {Kern},\ and\ \citenamefont {Ast}}]{Gierz2010}%
  \BibitemOpen
  \bibfield  {author} {\bibinfo {author} {\bibfnamefont {I.}~\bibnamefont
  {Gierz}}, \bibinfo {author} {\bibfnamefont {B.}~\bibnamefont
  {Stadtm\"uller}}, \bibinfo {author} {\bibfnamefont {J.}~\bibnamefont
  {Vuorinen}}, \bibinfo {author} {\bibfnamefont {M.}~\bibnamefont {Lindroos}},
  \bibinfo {author} {\bibfnamefont {F.}~\bibnamefont {Meier}}, \bibinfo
  {author} {\bibfnamefont {J.~H.}\ \bibnamefont {Dil}}, \bibinfo {author}
  {\bibfnamefont {K.}~\bibnamefont {Kern}}, \ and\ \bibinfo {author}
  {\bibfnamefont {C.~R.}\ \bibnamefont {Ast}},\ }\href {\doibase
  10.1103/PhysRevB.81.245430} {\bibfield  {journal} {\bibinfo  {journal} {Phys.
  Rev. B}\ }\textbf {\bibinfo {volume} {81}},\ \bibinfo {pages} {245430}
  (\bibinfo {year} {2010})}\BibitemShut {NoStop}%
\bibitem [{\citenamefont {Moreschini}\ \emph {et~al.}(2009)\citenamefont
  {Moreschini}, \citenamefont {Bendounan}, \citenamefont {Bentmann},
  \citenamefont {Assig}, \citenamefont {Kern}, \citenamefont {Reinert},
  \citenamefont {Henk}, \citenamefont {Ast},\ and\ \citenamefont
  {Grioni}}]{Moreschini2009}%
  \BibitemOpen
  \bibfield  {author} {\bibinfo {author} {\bibfnamefont {L.}~\bibnamefont
  {Moreschini}}, \bibinfo {author} {\bibfnamefont {A.}~\bibnamefont
  {Bendounan}}, \bibinfo {author} {\bibfnamefont {H.}~\bibnamefont {Bentmann}},
  \bibinfo {author} {\bibfnamefont {M.}~\bibnamefont {Assig}}, \bibinfo
  {author} {\bibfnamefont {K.}~\bibnamefont {Kern}}, \bibinfo {author}
  {\bibfnamefont {F.}~\bibnamefont {Reinert}}, \bibinfo {author} {\bibfnamefont
  {J.}~\bibnamefont {Henk}}, \bibinfo {author} {\bibfnamefont {C.~R.}\
  \bibnamefont {Ast}}, \ and\ \bibinfo {author} {\bibfnamefont
  {M.}~\bibnamefont {Grioni}},\ }\href {\doibase 10.1103/PhysRevB.80.035438}
  {\bibfield  {journal} {\bibinfo  {journal} {Phys. Rev. B}\ }\textbf {\bibinfo
  {volume} {80}},\ \bibinfo {pages} {035438} (\bibinfo {year}
  {2009})}\BibitemShut {NoStop}%
\bibitem [{\citenamefont {Bentmann}\ \emph {et~al.}(2009)\citenamefont
  {Bentmann}, \citenamefont {Forster}, \citenamefont {Bihlmayer}, \citenamefont
  {Chulkov}, \citenamefont {Moreschini}, \citenamefont {Grioni},\ and\
  \citenamefont {Reinert}}]{BentmannEPL2009}%
  \BibitemOpen
  \bibfield  {author} {\bibinfo {author} {\bibfnamefont {H.}~\bibnamefont
  {Bentmann}}, \bibinfo {author} {\bibfnamefont {F.}~\bibnamefont {Forster}},
  \bibinfo {author} {\bibfnamefont {G.}~\bibnamefont {Bihlmayer}}, \bibinfo
  {author} {\bibfnamefont {E.~V.}\ \bibnamefont {Chulkov}}, \bibinfo {author}
  {\bibfnamefont {L.}~\bibnamefont {Moreschini}}, \bibinfo {author}
  {\bibfnamefont {M.}~\bibnamefont {Grioni}}, \ and\ \bibinfo {author}
  {\bibfnamefont {F.}~\bibnamefont {Reinert}},\ }\href
  {http://stacks.iop.org/0295-5075/87/i=3/a=37003} {\bibfield  {journal}
  {\bibinfo  {journal} {EPL (Europhys. Lett.)}\ }\textbf {\bibinfo {volume}
  {87}},\ \bibinfo {pages} {37003} (\bibinfo {year} {2009})}\BibitemShut
  {NoStop}%
\bibitem [{\citenamefont {Bentmann}\ \emph {et~al.}(2011)\citenamefont
  {Bentmann}, \citenamefont {Kuzumaki}, \citenamefont {Bihlmayer},
  \citenamefont {Bl\"ugel}, \citenamefont {Chulkov}, \citenamefont {Reinert},\
  and\ \citenamefont {Sakamoto}}]{Bentmann2011}%
  \BibitemOpen
  \bibfield  {author} {\bibinfo {author} {\bibfnamefont {H.}~\bibnamefont
  {Bentmann}}, \bibinfo {author} {\bibfnamefont {T.}~\bibnamefont {Kuzumaki}},
  \bibinfo {author} {\bibfnamefont {G.}~\bibnamefont {Bihlmayer}}, \bibinfo
  {author} {\bibfnamefont {S.}~\bibnamefont {Bl\"ugel}}, \bibinfo {author}
  {\bibfnamefont {E.~V.}\ \bibnamefont {Chulkov}}, \bibinfo {author}
  {\bibfnamefont {F.}~\bibnamefont {Reinert}}, \ and\ \bibinfo {author}
  {\bibfnamefont {K.}~\bibnamefont {Sakamoto}},\ }\href {\doibase
  10.1103/PhysRevB.84.115426} {\bibfield  {journal} {\bibinfo  {journal} {Phys.
  Rev. B}\ }\textbf {\bibinfo {volume} {84}},\ \bibinfo {pages} {115426}
  (\bibinfo {year} {2011})}\BibitemShut {NoStop}%
\bibitem [{\citenamefont {\"Unal}\ \emph {et~al.}(2012)\citenamefont {\"Unal},
  \citenamefont {Winkelmann}, \citenamefont {Tusche}, \citenamefont {Bisio},
  \citenamefont {Ellguth}, \citenamefont {Chiang}, \citenamefont {Henk},\ and\
  \citenamefont {Kirschner}}]{Uenal2012}%
  \BibitemOpen
  \bibfield  {author} {\bibinfo {author} {\bibfnamefont {A.~A.}\ \bibnamefont
  {\"Unal}}, \bibinfo {author} {\bibfnamefont {A.}~\bibnamefont {Winkelmann}},
  \bibinfo {author} {\bibfnamefont {C.}~\bibnamefont {Tusche}}, \bibinfo
  {author} {\bibfnamefont {F.}~\bibnamefont {Bisio}}, \bibinfo {author}
  {\bibfnamefont {M.}~\bibnamefont {Ellguth}}, \bibinfo {author} {\bibfnamefont
  {C.-T.}\ \bibnamefont {Chiang}}, \bibinfo {author} {\bibfnamefont
  {J.}~\bibnamefont {Henk}}, \ and\ \bibinfo {author} {\bibfnamefont
  {J.}~\bibnamefont {Kirschner}},\ }\href {\doibase 10.1103/PhysRevB.86.125447}
  {\bibfield  {journal} {\bibinfo  {journal} {Phys. Rev. B}\ }\textbf {\bibinfo
  {volume} {86}},\ \bibinfo {pages} {125447} (\bibinfo {year}
  {2012})}\BibitemShut {NoStop}%
\bibitem [{\citenamefont {Ast}\ \emph {et~al.}(2007{\natexlab{b}})\citenamefont
  {Ast}, \citenamefont {Wittich}, \citenamefont {Wahl}, \citenamefont
  {Vogelgesang}, \citenamefont {Pacil\'e}, \citenamefont {Falub}, \citenamefont
  {Moreschini}, \citenamefont {Papagno}, \citenamefont {Grioni},\ and\
  \citenamefont {Kern}}]{Ast2007}%
  \BibitemOpen
  \bibfield  {author} {\bibinfo {author} {\bibfnamefont {C.~R.}\ \bibnamefont
  {Ast}}, \bibinfo {author} {\bibfnamefont {G.}~\bibnamefont {Wittich}},
  \bibinfo {author} {\bibfnamefont {P.}~\bibnamefont {Wahl}}, \bibinfo {author}
  {\bibfnamefont {R.}~\bibnamefont {Vogelgesang}}, \bibinfo {author}
  {\bibfnamefont {D.}~\bibnamefont {Pacil\'e}}, \bibinfo {author}
  {\bibfnamefont {M.~C.}\ \bibnamefont {Falub}}, \bibinfo {author}
  {\bibfnamefont {L.}~\bibnamefont {Moreschini}}, \bibinfo {author}
  {\bibfnamefont {M.}~\bibnamefont {Papagno}}, \bibinfo {author} {\bibfnamefont
  {M.}~\bibnamefont {Grioni}}, \ and\ \bibinfo {author} {\bibfnamefont
  {K.}~\bibnamefont {Kern}},\ }\href {\doibase 10.1103/PhysRevB.75.201401}
  {\bibfield  {journal} {\bibinfo  {journal} {Phys. Rev. B}\ }\textbf {\bibinfo
  {volume} {75}},\ \bibinfo {pages} {201401} (\bibinfo {year}
  {2007}{\natexlab{b}})}\BibitemShut {NoStop}%
\bibitem [{\citenamefont {Pascual}\ \emph {et~al.}(2004)\citenamefont
  {Pascual}, \citenamefont {Bihlmayer}, \citenamefont {Koroteev}, \citenamefont
  {Rust}, \citenamefont {Ceballos}, \citenamefont {Hansmann}, \citenamefont
  {Horn}, \citenamefont {Chulkov}, \citenamefont {Bl\"ugel}, \citenamefont
  {Echenique},\ and\ \citenamefont {Hofmann}}]{Pascual04}%
  \BibitemOpen
  \bibfield  {author} {\bibinfo {author} {\bibfnamefont {J.~I.}\ \bibnamefont
  {Pascual}}, \bibinfo {author} {\bibfnamefont {G.}~\bibnamefont {Bihlmayer}},
  \bibinfo {author} {\bibfnamefont {Y.~M.}\ \bibnamefont {Koroteev}}, \bibinfo
  {author} {\bibfnamefont {H.-P.}\ \bibnamefont {Rust}}, \bibinfo {author}
  {\bibfnamefont {G.}~\bibnamefont {Ceballos}}, \bibinfo {author}
  {\bibfnamefont {M.}~\bibnamefont {Hansmann}}, \bibinfo {author}
  {\bibfnamefont {K.}~\bibnamefont {Horn}}, \bibinfo {author} {\bibfnamefont
  {E.~V.}\ \bibnamefont {Chulkov}}, \bibinfo {author} {\bibfnamefont
  {S.}~\bibnamefont {Bl\"ugel}}, \bibinfo {author} {\bibfnamefont {P.~M.}\
  \bibnamefont {Echenique}}, \ and\ \bibinfo {author} {\bibfnamefont
  {P.}~\bibnamefont {Hofmann}},\ }\href {\doibase
  10.1103/PhysRevLett.93.196802} {\bibfield  {journal} {\bibinfo  {journal}
  {Phys. Rev. Lett.}\ }\textbf {\bibinfo {volume} {93}},\ \bibinfo {pages}
  {196802} (\bibinfo {year} {2004})}\BibitemShut {NoStop}%
\bibitem [{\citenamefont {Roushan}\ \emph {et~al.}(2009)\citenamefont
  {Roushan}, \citenamefont {Seo}, \citenamefont {Parker}, \citenamefont {Hor},
  \citenamefont {Hsieh}, \citenamefont {Qian}, \citenamefont {Richardella},
  \citenamefont {Hasan}, \citenamefont {Cava},\ and\ \citenamefont
  {Yazdani}}]{roushan_topological_2009}%
  \BibitemOpen
  \bibfield  {author} {\bibinfo {author} {\bibfnamefont {P.}~\bibnamefont
  {Roushan}}, \bibinfo {author} {\bibfnamefont {J.}~\bibnamefont {Seo}},
  \bibinfo {author} {\bibfnamefont {C.~V.}\ \bibnamefont {Parker}}, \bibinfo
  {author} {\bibfnamefont {Y.~S.}\ \bibnamefont {Hor}}, \bibinfo {author}
  {\bibfnamefont {D.}~\bibnamefont {Hsieh}}, \bibinfo {author} {\bibfnamefont
  {D.}~\bibnamefont {Qian}}, \bibinfo {author} {\bibfnamefont {A.}~\bibnamefont
  {Richardella}}, \bibinfo {author} {\bibfnamefont {M.~Z.}\ \bibnamefont
  {Hasan}}, \bibinfo {author} {\bibfnamefont {R.~J.}\ \bibnamefont {Cava}}, \
  and\ \bibinfo {author} {\bibfnamefont {A.}~\bibnamefont {Yazdani}},\ }\href
  {\doibase 10.1038/nature08308} {\bibfield  {journal} {\bibinfo  {journal}
  {Nature}\ }\textbf {\bibinfo {volume} {460}},\ \bibinfo {pages} {1106}
  (\bibinfo {year} {2009})}\BibitemShut {NoStop}%
\bibitem [{\citenamefont {Walls}\ and\ \citenamefont
  {Heller}(2007)}]{walls_spin-orbit_2007}%
  \BibitemOpen
  \bibfield  {author} {\bibinfo {author} {\bibfnamefont {J.~D.}\ \bibnamefont
  {Walls}}\ and\ \bibinfo {author} {\bibfnamefont {E.~J.}\ \bibnamefont
  {Heller}},\ }\href {\doibase 10.1021/nl071711z} {\bibfield  {journal}
  {\bibinfo  {journal} {Nano Lett.}\ }\textbf {\bibinfo {volume} {7}},\
  \bibinfo {pages} {3377} (\bibinfo {year} {2007})}\BibitemShut {NoStop}%
\bibitem [{\citenamefont {Kaminski}\ \emph {et~al.}(2005)\citenamefont
  {Kaminski}, \citenamefont {Poodt}, \citenamefont {Aret}, \citenamefont
  {Radenovic},\ and\ \citenamefont {Vlieg}}]{Kaminski2005}%
  \BibitemOpen
  \bibfield  {author} {\bibinfo {author} {\bibfnamefont {D.}~\bibnamefont
  {Kaminski}}, \bibinfo {author} {\bibfnamefont {P.}~\bibnamefont {Poodt}},
  \bibinfo {author} {\bibfnamefont {E.}~\bibnamefont {Aret}}, \bibinfo {author}
  {\bibfnamefont {N.}~\bibnamefont {Radenovic}}, \ and\ \bibinfo {author}
  {\bibfnamefont {E.}~\bibnamefont {Vlieg}},\ }\href {\doibase
  10.1016/j.susc.2004.11.001} {\bibfield  {journal} {\bibinfo  {journal} {Surf.
  Sci.}\ }\textbf {\bibinfo {volume} {575}},\ \bibinfo {pages} {233 } (\bibinfo
  {year} {2005})}\BibitemShut {NoStop}%
\bibitem [{\citenamefont {Crommie}\ \emph {et~al.}(1993)\citenamefont
  {Crommie}, \citenamefont {Lutz},\ and\ \citenamefont {Eigler}}]{Crommie1993}%
  \BibitemOpen
  \bibfield  {author} {\bibinfo {author} {\bibfnamefont {M.~F.}\ \bibnamefont
  {Crommie}}, \bibinfo {author} {\bibfnamefont {C.~P.}\ \bibnamefont {Lutz}}, \
  and\ \bibinfo {author} {\bibfnamefont {D.~M.}\ \bibnamefont {Eigler}},\
  }\href {\doibase 10.1038/363524a0} {\bibfield  {journal} {\bibinfo  {journal}
  {Nature}\ }\textbf {\bibinfo {volume} {363}},\ \bibinfo {pages} {524}
  (\bibinfo {year} {1993})}\BibitemShut {NoStop}%
\bibitem [{\citenamefont {Hasegawa}\ and\ \citenamefont
  {Avouris}(1993)}]{Hasegawa1993}%
  \BibitemOpen
  \bibfield  {author} {\bibinfo {author} {\bibfnamefont {Y.}~\bibnamefont
  {Hasegawa}}\ and\ \bibinfo {author} {\bibfnamefont {P.}~\bibnamefont
  {Avouris}},\ }\href {\doibase 10.1103/PhysRevLett.71.1071} {\bibfield
  {journal} {\bibinfo  {journal} {Phys. Rev. Lett.}\ }\textbf {\bibinfo
  {volume} {71}},\ \bibinfo {pages} {1071} (\bibinfo {year}
  {1993})}\BibitemShut {NoStop}%
\bibitem [{\citenamefont {Sprunger}\ \emph {et~al.}(1997)\citenamefont
  {Sprunger}, \citenamefont {Petersen}, \citenamefont {Plummer}, \citenamefont
  {L{\ae}gsgaard},\ and\ \citenamefont {Besenbacher}}]{Sprunger1997}%
  \BibitemOpen
  \bibfield  {author} {\bibinfo {author} {\bibfnamefont {P.~T.}\ \bibnamefont
  {Sprunger}}, \bibinfo {author} {\bibfnamefont {L.}~\bibnamefont {Petersen}},
  \bibinfo {author} {\bibfnamefont {E.~W.}\ \bibnamefont {Plummer}}, \bibinfo
  {author} {\bibfnamefont {E.}~\bibnamefont {L{\ae}gsgaard}}, \ and\ \bibinfo
  {author} {\bibfnamefont {F.}~\bibnamefont {Besenbacher}},\ }\href {\doibase
  10.1126/science.275.5307.1764} {\bibfield  {journal} {\bibinfo  {journal}
  {Science}\ }\textbf {\bibinfo {volume} {275}},\ \bibinfo {pages} {1764}
  (\bibinfo {year} {1997})}\BibitemShut {NoStop}%
\bibitem [{Note1()}]{Note1}%
  \BibitemOpen
  \bibinfo {note} {See Supplemental Material at [URL] for a movie of $dI/dV$
  maps between -0.45~V and +0.4~V.}\BibitemShut {Stop}%
\bibitem [{Note2()}]{Note2}%
  \BibitemOpen
  \bibinfo {note} {Determining $q_1$ and $q_{12}$ close to the band maximum
  from FT cross sections becomes difficult; here our analysis is based on a
  combined FT and $dI/dV$-linescan analysis.}\BibitemShut {Stop}%
\bibitem [{\citenamefont {Hirayama}\ \emph {et~al.}(2011)\citenamefont
  {Hirayama}, \citenamefont {Aoki},\ and\ \citenamefont {Kato}}]{Hirayama2011}%
  \BibitemOpen
  \bibfield  {author} {\bibinfo {author} {\bibfnamefont {H.}~\bibnamefont
  {Hirayama}}, \bibinfo {author} {\bibfnamefont {Y.}~\bibnamefont {Aoki}}, \
  and\ \bibinfo {author} {\bibfnamefont {C.}~\bibnamefont {Kato}},\ }\href
  {\doibase 10.1103/PhysRevLett.107.027204} {\bibfield  {journal} {\bibinfo
  {journal} {Phys. Rev. Lett.}\ }\textbf {\bibinfo {volume} {107}},\ \bibinfo
  {pages} {027204} (\bibinfo {year} {2011})}\BibitemShut {NoStop}%
\bibitem [{\citenamefont {Pascual}\ \emph {et~al.}(2001)\citenamefont
  {Pascual}, \citenamefont {Song}, \citenamefont {Jackiw}, \citenamefont
  {Horn},\ and\ \citenamefont {Rust}}]{Pascual2001}%
  \BibitemOpen
  \bibfield  {author} {\bibinfo {author} {\bibfnamefont {J.~I.}\ \bibnamefont
  {Pascual}}, \bibinfo {author} {\bibfnamefont {Z.}~\bibnamefont {Song}},
  \bibinfo {author} {\bibfnamefont {J.~J.}\ \bibnamefont {Jackiw}}, \bibinfo
  {author} {\bibfnamefont {K.}~\bibnamefont {Horn}}, \ and\ \bibinfo {author}
  {\bibfnamefont {H.-P.}\ \bibnamefont {Rust}},\ }\href {\doibase
  10.1103/PhysRevB.63.241103} {\bibfield  {journal} {\bibinfo  {journal} {Phys.
  Rev. B}\ }\textbf {\bibinfo {volume} {63}},\ \bibinfo {pages} {241103}
  (\bibinfo {year} {2001})}\BibitemShut {NoStop}%
\bibitem [{\citenamefont {Schouteden}\ \emph {et~al.}(2009)\citenamefont
  {Schouteden}, \citenamefont {Lievens},\ and\ \citenamefont
  {Van~Haesendonck}}]{Schouteden2009}%
  \BibitemOpen
  \bibfield  {author} {\bibinfo {author} {\bibfnamefont {K.}~\bibnamefont
  {Schouteden}}, \bibinfo {author} {\bibfnamefont {P.}~\bibnamefont {Lievens}},
  \ and\ \bibinfo {author} {\bibfnamefont {C.}~\bibnamefont
  {Van~Haesendonck}},\ }\href {\doibase 10.1103/PhysRevB.79.195409} {\bibfield
  {journal} {\bibinfo  {journal} {Phys. Rev. B}\ }\textbf {\bibinfo {volume}
  {79}},\ \bibinfo {pages} {195409} (\bibinfo {year} {2009})}\BibitemShut
  {NoStop}%
\bibitem [{\citenamefont {Mirhosseini}\ \emph {et~al.}(2009)\citenamefont
  {Mirhosseini}, \citenamefont {Henk}, \citenamefont {Ernst}, \citenamefont
  {Ostanin}, \citenamefont {Chiang}, \citenamefont {Yu}, \citenamefont
  {Winkelmann},\ and\ \citenamefont {Kirschner}}]{Mirhosseini2009}%
  \BibitemOpen
  \bibfield  {author} {\bibinfo {author} {\bibfnamefont {H.}~\bibnamefont
  {Mirhosseini}}, \bibinfo {author} {\bibfnamefont {J.}~\bibnamefont {Henk}},
  \bibinfo {author} {\bibfnamefont {A.}~\bibnamefont {Ernst}}, \bibinfo
  {author} {\bibfnamefont {S.}~\bibnamefont {Ostanin}}, \bibinfo {author}
  {\bibfnamefont {C.-T.}\ \bibnamefont {Chiang}}, \bibinfo {author}
  {\bibfnamefont {P.}~\bibnamefont {Yu}}, \bibinfo {author} {\bibfnamefont
  {A.}~\bibnamefont {Winkelmann}}, \ and\ \bibinfo {author} {\bibfnamefont
  {J.}~\bibnamefont {Kirschner}},\ }\href {\doibase 10.1103/PhysRevB.79.245428}
  {\bibfield  {journal} {\bibinfo  {journal} {Phys. Rev. B}\ }\textbf {\bibinfo
  {volume} {79}},\ \bibinfo {pages} {245428} (\bibinfo {year}
  {2009})}\BibitemShut {NoStop}%
\bibitem [{\citenamefont {Wegner}\ \emph {et~al.}(2006)\citenamefont {Wegner},
  \citenamefont {Bauer}, \citenamefont {Koroteev}, \citenamefont {Bihlmayer},
  \citenamefont {Chulkov}, \citenamefont {Echenique},\ and\ \citenamefont
  {Kaindl}}]{Wegner2006}%
  \BibitemOpen
  \bibfield  {author} {\bibinfo {author} {\bibfnamefont {D.}~\bibnamefont
  {Wegner}}, \bibinfo {author} {\bibfnamefont {A.}~\bibnamefont {Bauer}},
  \bibinfo {author} {\bibfnamefont {Y.~M.}\ \bibnamefont {Koroteev}}, \bibinfo
  {author} {\bibfnamefont {G.}~\bibnamefont {Bihlmayer}}, \bibinfo {author}
  {\bibfnamefont {E.~V.}\ \bibnamefont {Chulkov}}, \bibinfo {author}
  {\bibfnamefont {P.~M.}\ \bibnamefont {Echenique}}, \ and\ \bibinfo {author}
  {\bibfnamefont {G.}~\bibnamefont {Kaindl}},\ }\href@noop {} {\bibfield
  {journal} {\bibinfo  {journal} {Phys. Rev. B}\ }\textbf {\bibinfo {volume}
  {73}},\ \bibinfo {pages} {115403} (\bibinfo {year} {2006})}\BibitemShut
  {NoStop}%
\bibitem [{\citenamefont {Kr\"uger}()}]{Krueger}%
  \BibitemOpen
  \bibfield  {author} {\bibinfo {author} {\bibfnamefont {P.}~\bibnamefont
  {Kr\"uger}},\ }\href@noop {} {}\bibinfo {note} {Private
  communication}\BibitemShut {NoStop}%
\end{thebibliography}%

\end{document}